\documentclass[epj]{svjour}

\usepackage{graphics}
\begin{document}

\title{A causal multifractal stochastic equation and its statistical properties}
\author{ Fran\c{c}ois G. Schmitt  % etc
}                     % Do not remove
\institute{CNRS, UMR ELICO, Station Marine, Universit\'e de Lille 1, BP 80,
62930 Wimereux, France, \email{francois.schmitt@univ-lille1.fr}}
\date{Received: date / Revised version: date}
\abstract{
Multiplicative cascades have been introduced in
turbulence to generate random or deterministic fields
having intermittent values and long-range power-law
correlations. Generally this is done using discrete
construction rules leading to discrete cascades. Here
a causal log-normal stochastic process is introduced; its 
multifractal properties are demonstrated together
with other properties such as the composition rule for
scale dependence and stochastic differential equations
for time and scale evolutions. This multifractal stochastic process
is continuous in scale ratio and in time. It has a simple
generating equation and can be used to generate
sequentially time series of any length.
 % end of PACS codes
} %end of abstract
\maketitle
\section{Introduction}
\label{intro}
Since their introduction in the fields
of turbulence, geophysics and chaos theory, multifractals
have been widely used in many fields of applied
sciences, including turbulence, geophysics, precipitations,
oceanography, high energy physics, astronomy, biology and finance.
The main characteristics of multifractal models or data is
to possess high variability on a wide range of time or
space scales, associated to intermittent fluctuations
and long-range power-law correlations.
 In this framework, many studies have been devoted to
data analysis and discrete mathematical model constructions.
More precisely, there is a dichotomy in the multifractal
literature between, on the one hand, experimental results
showing usually no characteristic scales in a scaling
range between an outer and an inner scale, and on the other
hand, discrete cascade models that are not stationary and
are scaling only for discrete scales ratios. This is why continuous
multifractal models are needed, and among them, multifractal
stochastic processes: processes possessing multifractal statistics while 
depending on a continuous time parameter.
Here we consider one such process, belonging to the generatic family
of lognormal multifractals, while depending on a continuous time
parameter. This process has a simple analytical expression, and is
causal and 1D. We will discuss some of its properties, and confirm them using
data analysis of a numerical simulation. 

Below we first recall the basic framework of discrete cascades:
their statistics and correlation properties. These properties
are given as the definition of a multifractal
cascade process, and are recovered for continuous causal
processes. In section 2, we discuss the continuous scale limit of
discrete cascades, and provide explicitly a basic causal equation generating
a multifractal stochastic process for log-normal multifractals.
We explore their statistical properties, including correlations
for the process and its logarithm, and provide differential 
forms in time and in scale space. A link with the Langevin equation
for Markov cascades is presented. Section 3 presents a simulation of a discrete
cascade and a time simulation for
a multifractal continuous process, together with a data analysis which confirms the
theoretical results for both simulations. Section 4 is 
devoted to concluding
remarks. Some useful results in probability theory
are recalled in the Appendices.

\subsection{The generic example of multiplicative cascades in
turbulence}

We briefly recall here the classical framework of Richardson turbulent 
cascades, which can be considered as the generic
example of cascades.
For high Reynolds number turbulence,
dissipative terms are negligible compared to
the inertial terms of the Navier-Stokes equations,
so that for the inertial range, there is a local 
energy transfer from large to small scales. This 
is the picture of Richardson's 1922 energy cascade \cite{rich22},
which was later formalized by Kolmogorov in 1941 \cite{kolm41},
corresponding to the Komogorov-Obukhov -5/3 Fourier power spectrum for
velocity fluctuations in the inertial range.
This description has been found experimentally valid for mean fluctuations,
and this general cascade framework for the inertial range is
now well accepted. Nevertheless it was not able to take into
account intermittent fluctuations of the small-scale dissipation
field, as experimentally observed a few years later
\cite{batc49}. This intermittency lead Kolmogorov and
Obukhov to propose new models in 1962 \cite{kolm62,obuk62},
assuming log-normal statistics for the velocity increments
and for the dissipation field $\epsilon$, with furthermore the
following asymptotic hypothesis for the variance of $\log \epsilon$:
\begin{equation}
\sigma^2_{\log \epsilon} \sim A + \mu \log \lambda
\label{eq0}
\end{equation}
where $A$ and $\mu>0$ are constants
($\mu$ is often called the intermittency exponent), and $\lambda=L/\ell$
is the scale ratio of the outer scale $L$ to the scale of 
interest $\ell$ (in the same paper this property of the variance was also assumed for the 
averaged dissipation $\epsilon_r$ over a sphere of radius $r$). 
Kolmogorov's proposals in 1962 were given without
any physical or modelling justification; it was rather ad hoc.
Soon after this, several experimental studies have shown that
the dissipation field possesses long-range power-law
correlations \cite{gurv63,pond65}. This was a very specific property which
was not included in the proposal of Kolmogorov. 
This lead Yaglom to propose in 1966 his explicit cascade
model, which was able to reproduce experimental facts --
small-scale intermittency and power-law correlations --
and the log-normal hypothesis of Kolmogorov. This multiplicative
cascade model had hence experimental grounds, and was proposing
an unification of different results and proposals. It is discussed
below.

\subsection{Discrete multiplicative cascades and their properties}

We present now discrete multiplicative cascades and their
statistical properties. 
This is basically a discrete model in scale, but it
can be densified. The terms ``discrete in scale'' refer
to the fact that the scale ratio from mother to daughter
structures is an integer number strictly larger than 1.
This model is multiplicative, and embedded in a recursive manner.
The multiplicative hypothesis generates large fluctuations,
and the embedding generates long-range correlations, that give
to these large fluctuations their intermittent character.

\begin{figure}
\begin{center}
\resizebox{0.40\textwidth}{!}{%  
  \includegraphics{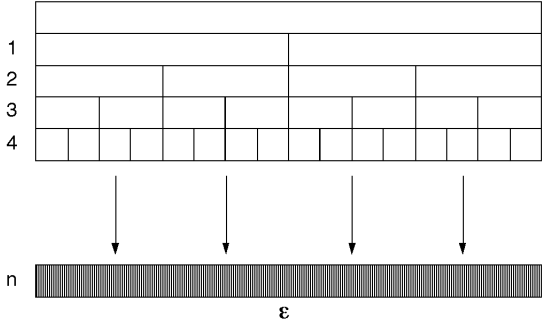}
}
\end{center}
%\vspace{5cm}       % Give the correct figure height in cm
\caption{Illustration of the discrete cascade process. 
Each step is associated to a scale ratio of 2. 
After n steps, the total scale ratio is $2^n$.}
\label{fig:1}       % Give a unique label
\end{figure}

\subsubsection{Scaling properties of discrete multiplicative cascades}

We symbolize the eddy cascade by cells, each cell being
associated to a random variable $W_i$, called ``weight''. 
All these random variables are assumed positive, independent, and
identically distributed, with the following conservative
property: $<W>=1$, where $<.>$ denotes expectation.
The larger scale corresponds to a unique cell of size 
$L=\ell_0 \lambda_1^n$, where $\ell_0$ is a fixed scale and
$\lambda_1>1$ is an adimensional scale ratio (very often for
discrete models one takes $\lambda_1 = 2$). The model being discrete,
the next scale involved corresponds to $\lambda_1$
cells, each of size $L/\lambda_1=\ell_0 \lambda_1^{n-1}$.
This is iterated and at step $p$ ($1 \le p \le n$) there
are $\lambda_1^p$ cells, each of size $L/\lambda_1^p=\ell_0 \lambda_1^{n-p}$.
There are $n$ cascade steps, and at step $n$ there are $\lambda_1^n$
cells, each of size $\ell_0$, which is the smallest scale of the cascade
(see Figure 1). To reach this scale, in agreement with
Richardson's original idea \cite{rich22} and also with more recent
results on the localness of interactions in turbulence \cite{fris95},
all intermediate scales have been involved. Finally, at each point the
energy flux writes as the product of $n$ weights \cite{yagl66}:
\begin{equation}
\epsilon(x) = \prod_{p=1}^n W_{p,x}
\label{eq1}
\end{equation}
\noindent where $W_{p,x}$ is the random variable corresponding to
position $x$ and level $p$ in the cascade. Since each $W_{p,x}$ for
different cells are assumed independent, their moment of order 
$q>0$ can be estimated as:
\begin{equation}
 < (\epsilon(x))^q > =  \prod_{i=1}^n < \left( W_{p,x}\right)^q > 
 = < W^q >^n
\label{eq2}
\end{equation}
Introducing the total scale ratio
\begin{equation}
    \lambda = \frac{L}{\ell_0} = \lambda_1^n
\label{eq3}
\end{equation}
between the outer and the inner scales, Eq. \ref{eq2}
gives finally the moments of the small-scale flux,
denoted hereafter $\epsilon_{\lambda}(x)$ to indicate its scale-ratio
dependence
\begin{equation}
 < \epsilon_{\lambda}^q > =   \lambda^{K(q)}
\label{eq4}
\end{equation}
where we introduced the cumulant generating function 
$K(q) = \log_{\lambda_1} < W^q >$. The conservative property
$<W>=1$ gives $K(1)=0$ and also $<\epsilon>=1$. 

\indent Equation \ref{eq4} is a generic property of multifractal fields,
obtained through multiplicative cascades. Depending on the model
chosen, different forms of the pdf of $W$ can be
taken, leading to different analytical expressions for $K(q)$.
A log-normal pdf for $W$ corresponds to a quadratic expression
$K(q)=\frac{\mu}{2}(q^2-q)$, with $\mu=K(2)=\sigma^2/\log2$,
where $\sigma^2$ is the variance of $\log W$ (this is detailed
below).
It can be also noticed that $K(q)$ is (up to a $\log \lambda_1$
factor) the second Laplace characteristic function of the random
variable $\log W$ (see the Appendix A), showing that it is a
convex function (see \cite{fell71}).

Let us mention here a specific property of multiplicative
cascades: up to now, we have considered a cascading 
process from large to small scales, called ``bare'' field
by Mandelbrot \cite{mand74}, corresponding to the energy flux
in turbulence. When the smallest scale is reached, the Kolmogorov
dissipation scale $\eta$, the flux becomes the dissipation.
An interesting question is here to characterize
the ``dressed'' field, corresponding to integrated small-scale dissipation
as proposed by Kolmogorov and Obukhov \cite{kolm62}.
Dressed quantities (e.g. observables) are a measure, up to a
resolution scale $r \gg \eta$, of the small-scale cascade, 
developed down to scale $\eta$. In fact, the dressing may
introduce divergence of moments \cite{mand74,sche87} but it is known
that for moments below this critical moment $q_D$, the scaling exponents
of bare and dressed fields are identical. An interesting mathematical
question is also the convergence of the measure when the total scale
ratio $\lambda=L/\eta \rightarrow \infty$ \cite{kaha76}. But this point is not the 
main topic of our paper. Indeed, the dressing is usually seen as performed
by the observer; we therefore consider here our stochastic process (presented below)
as representative of the cascade developed from a large to a smaller scale,
i.e. a bare cascade. The question of the interrelation of dressed-bare fields
and their respective statistical moments is therefore the same for our
stochastic process, as for all multifractal cascade models (discrete or
continuous).

\begin{figure}
\begin{center}
\resizebox{0.40\textwidth}{!}{%  
  \includegraphics{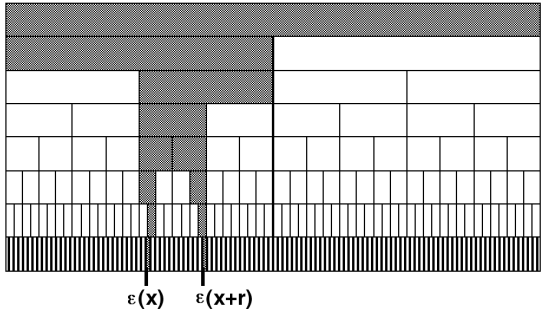}
}
\end{center}
\caption{The paths leading to 2 points separated by a 
distance $r$: they have a common path before a bifurcation.}
\label{fig:2}       % Give a unique label
\end{figure}

\subsubsection{Correlation properties}

We now consider the long-range correlations generated by this cascade
model. As originally shown by Yaglom \cite{yagl66}, the correlation
$<\epsilon_{\lambda}(x)\epsilon_{\lambda}(x+r)>$ can be
decomposed as a product of random variables, introducing
their common ancestor as follows. The distance $r$ corresponds
in average to $m$ steps, with $\lambda_1^m  \approx r$, so that
the $n-m$ first steps are common to the two random variables, whereas
there is a ``bifurcation'' at the step $m$, and the two paths separate
(see Figure 2). In the product of the two random variables, there are
thus $n-m$ identical variables and $m$ different and independent.
This gives:
\begin{eqnarray}
&&  < \epsilon_{\lambda}(x)  \epsilon_{\lambda}(x+r)>  =  <\prod_{p=1}^n W_{p,x}
      \prod_{p'=1}^n W_{p',x+r}> \nonumber \\
      & = & \prod_{p=1}^{n-m} < W_{p,x}^2>\prod_{p=n-m+1}^n <W_{p,x}> \times \nonumber\\
      && \times \prod_{p'=n-m+1}^n <W_{p',x+r}> \nonumber\\
      & = & <W^2>^{n-m}<W>^{2m}
\label{eq5}
\end{eqnarray}
Finally, introducing $K(q)$ and recalling that $\lambda_1^n=\lambda$
and $\lambda_1^m \approx r$ this gives:
\begin{equation}
 <\epsilon_{\lambda}(x)\epsilon_{\lambda}(x+r)> \approx \lambda^{K(2)}r^{-K(2)}
\label{eq6}
\end{equation}
Since $\lambda$ is fixed, this relation corresponds to a long-range power-law 
correlation with exponent $\mu=K(2)$. Let us note that the condition $\lambda_1^m \approx r$ 
imposes that the scaling relation (\ref{eq6}) is exact for discrete scales only. This
is an important limitation of discrete cascades, together with the fact that they are
not strictly stationary.
Relation (\ref{eq6}) can be generalized to provide, following the same line,  the two-points correlations of 
moments of order $p>0$ and $q>0$
\cite{cate87,mars96}:
\begin{equation}
 <\epsilon_{\lambda}(x)^p \epsilon_{\lambda}(x+r)^q> \approx
  \lambda^{K(p+q)}  
r^{-(K(p+q)-K(p)-K(q))}
\label{eq6b}
\end{equation}
Since $K(q)$ is non-linear and convex for multifractal
distributions, the exponent $K(p+q)-K(p)-K(q) >0$ quantifies the long-range power law correlations
of multifractal measures. For $p=q=1$, this yields the
$\mu=K(2)$ exponent given above.

\indent On the other hand, we may also consider the correlation properties
of the logarithm of the process (called generator), which possesses also interesting
characteristics. Let us define the generators $\gamma$ and $g$ 
of respectively the cascade process $\epsilon$ and the weight $W$:
\begin{equation}
  \left\lbrace
    \begin{array}{l}
      \gamma = \log \epsilon \\
      g = \log W
     \end{array}
   \right.
\label{eq7}
\end{equation}
We consider the autocorrelation function of $\gamma$:
\begin{equation}
  C_{\gamma}(r) = <\gamma(x)\gamma(x+r)>
\label{eq8}
\end{equation}
As before, the path leading to $\gamma(x)$ and
$\gamma(x+r)$ may be split: for the last $m$ steps
the paths are different, whereas before this bifurcation, the paths
are common. The main difference with the calculation leading to
Eq.(\ref{eq6}) is that in Eq. (\ref{eq8}) $\gamma$ is not the
product but the sum of random variables. A simple calculation
shows that $C_{\gamma}(r)$ involves $n^2-n+m$ terms of the form
$<g>^2$ and $n-m$ terms of the form $<g^2>$. Let us note $G=<g>=<\log W>$
and $\sigma^2 = <(g-G)^2>=<g^2>-G^2$. We then have:
\begin{eqnarray}
   C_{\gamma}(r) & \approx & <g>^2(n^2-n+m)+<g^2>(n-m) \nonumber \\
        & \approx & G^2n^2 + \sigma^2 (n-m)   \nonumber \\
	& \approx & C - \frac{\sigma^2}{\log \lambda_1} \log r
\label{eq9}
\end{eqnarray}
where we introduced the constant $C=G^2n^2+\sigma^2n$. This
corresponds to a logarithmic decay of the autocorrelation function
of the generator \cite{arne98a,arne98b}. Its Fourier transform gives
the power spectrum of the singularity process $\gamma(x)$:
\begin{equation}
   E_{\gamma}(k) \approx k^{-1}
\label{eq10}
\end{equation}
corresponding to an exactly $1/f$ noise \cite{sche87,sche91,muzy00}.
Properties (\ref{eq4}), (\ref{eq6b}), (\ref{eq9}) and (\ref{eq10})
may be used to check the multifractality of a stochastic process. 
Analogous properties are verified for dressed multifractals and can hence
be used to test numerical simulations or experimental data
(for a link between log-correlations and dressed scaling properties,
see also \cite{muzy00,bacr01}).

\subsubsection{The log-normal case}

Up to now, the analytical
expression taken by $K(q)$ is only loosely constrained a priori:
a convex function verifying  $K(0)=0$ and $K(1)=0$.
We explicit here the generic example of multifractal cascade,
given by the log-normal case, where generators are Gaussian.
We choose here for simplicity $\lambda_1 = 2$. We take $W=e^g$,
where $g$ is a Gaussian random variable of mean $G$ and
variance $\sigma^2$: $g=\sigma g_0+G$, where $g_0$ is a centered and unitary Gaussian
random variable. We have:
\begin{equation}
  <W^q>= \exp \left( \frac{1}{2}\sigma^2q^2+qG \right)
\label{eq11}
\end{equation}
The conditions $<W>=1$ and $\mu=K(2)=\log_2<W^2>$ provide the constants,
and finally the weights of the form:
\begin{equation}
   W = \exp \left( \sqrt{\mu \log2} g_0 -\frac{\mu}{2}\log 2 \right)
\label{eq12}
\end{equation}
generate a discrete log-normal multifractal field with moment function
\begin{equation}
   K(q) = \frac{\mu}{2}\left( q^2 - q \right)
\label{eq13}
\end{equation}
This corresponds to log-normal statistics for $\epsilon$,
with a variance of $\log \epsilon$ obeying Kolmogorov's initial
hypothesis, given by Eq.(\ref{eq0}). Formula (\ref{eq12}) can be directly
used for discrete log-normal simulations.

\section{A causal log-normal stochastic evolution equation and its
time and scale properties}

\subsection{Introduction}
Up to now, a discrete cascade procedure has been presented, 
and its characteristic properties given. In this section we are interested 
in the scale densification, introducing more and more cascade steps, while 
maintaining constant the total scale ratio. We first recall below how this 
leads to log-infinitely divisible (log-ID) probability distributions. This considerably restricts 
possible pdf for continuous scale multifractal cascades. On the other hand, it is also 
very useful to have a stochastic equation generating continuous scale 
multifractal field or process at our disposal. Such a stochastic equation would depend on some 
parameters and a scale variable for fields, or on a time variable for
processes.  A few recent papers have been devoted to such 
topic, and we briefly present them below, before providing a stochastic equation 
corresponding to a causal log-normal multifractal process. We then study this 
stochastic equation: we show that this produces indeed a multifractal 
log-normal process with the desired correlation properties, and we discuss
further its differential form, when differentiated versus time and versus
scale variables.

\subsubsection{Continuous cascades lead to log-infinitely divisible
processes and stochastic measures}
To densify the cascade described above,
we keep the total scale ratio $\lambda$ large but fixed;
the continuous limit can be obtained by increasing the total step
number $n$, hence $\lambda_1 = \lambda^{1/n} \rightarrow 1^+$
(see \cite{sche91,she95,sche97,schm98}).
It is then well-known that, in this limit, 
$\log \epsilon$ is an ID random variable
(see Appendix A and \cite{fell71} for ID random variables):
continuous cascade models
are log-ID \cite{she95,sait92,novi94,cast96}. 
This restricts the eligible
cascade models, since ID laws define a specific family of
probability distributions.

In order to provide stochastic equations and processes 
generating log-ID continuous multifractals,  we have recently introduced 
ID stochastic integrals based on ID measures \cite{schm01a,schm01b}
(see also \cite{muzy02,chai03}). 
ID measures are set functions, also called in the mathematical 
literature independently scattered random measures \cite{rajp89}.
This provided general expressions for the whole log-ID
family. A more restricted but still quite general family is composed of
log-stable processes, for which generating equations take 
simpler forms \cite{sche87,mars96,schm01b}. Among these, 
we consider here a special but important case: log-normal processes. 
We study this here because the stochastic integral involved 
is the classical Gaussian stochastic integral, whose statistical and
differential properties are well-known; furthermore, log-normal multifractals
are classical (and historical) examples of multifractals. The results given here for
this type of continuous cascade may be generalized later for other 
ID processes. 

Let us underline that the bare/dressed question mentioned above for
discrete cascades is still valid for continuous cascades;
the convergence properties of dressed log-ID 
measures has been studied in \cite{muzy02,barr03}.

\subsubsection{The stochastic equation}
Until now, cascades have been presented in space; for the remaining of the
paper, we consider cascades developing in time. The cascade is then
seen as a stochastic process possessing long-range correlations for
time increments $\tau$.

The stochastic equation we study here is the following:
\begin{equation}
  \epsilon_{\lambda}(t) = \lambda^{-\mu/2} \exp \left(
    \mu^{1/2}\int_{t+1-\lambda}^t (t+1-u)^{-1/2} dB(u) \right)
\label{eq14}
\end{equation}
where $c>0$ is a constant, $\lambda \gg 1$ is the total
scale ratio, $t$ is time and $B(t)$ is a Brownian motion.
This was already given in \cite{schm01b} and in other less explicit
forms in \cite{sche87,sche91,mars96} but its statistical and
differential properties have not been studied in details in these
papers.
This equation corresponds to the exponential of a Gaussian stochastic integral,
this integral being analogous to a fractional integration of a Gaussian process,
i.e. a fractional Brownian motion \cite{mand68}.
The only difference with a formal fractional integration comes from the fact
that the domain of integration is finite (for large values of 
$\lambda$ it is a wide domain, but still finite). The exponent of this fractional
Brownian motion is not a free parameter. Let us recall that for a 
self-similar parameter $H$ the kernel in the integration for a fractional 
Brownian motion is on the form
$(t-u)^{H-1/2}$, so that the kernel introduced above corresponds to a
self-similar parameter $H=0$. This is thus a very special case of
fractional Brownian motion.
This equation is also causal (the past does not depend on the future)
and can be used to generate sequentially a log-normal process with no upper 
limit in the number of points. We study below in details the properties of this equation.

\indent It is first easily verified that this process is stationary:
\begin{equation}
  \epsilon_{\lambda}(t+t') \stackrel{d}{=} \epsilon_{\lambda}(t)
\label{eq15}
\end{equation}
where $\stackrel{d}{=}$ means equality in distribution.
The statistical moments are then independent of time, and
are estimated below:
\begin{eqnarray}
  < \epsilon_{\lambda}(t)^q >  &= & \lambda^{-q\mu/2}
     < e^{ q\mu^{1/2} \int_{t+1-\lambda}^t (t+1-u)^{-1/2} dB(u)}>
     \nonumber \\
     & = & \lambda^{-q\mu/2} 
     \exp \left( \frac{\mu q^2}{2}\int_{t+1-\lambda}^t \frac{du}{t+1-u} \right)
     \nonumber \\
     & = & \lambda^{-q\mu/2} \lambda^{\frac{\mu q^2}{2}}
\label{eq16}
\end{eqnarray}
giving the requested scale-invariant properties, Eq.(\ref{eq4}) and Eq. (\ref{eq13}) (see Appendix B for
the second characteristic function of a stable stochastic integral). We explore below 
other properties of this equation in time and in scale-space.

\subsection{Time statistics of the log-normal stochastic process}

\subsubsection{Correlations}
We consider here the generalized correlation function
\begin{equation} 
  C_{\epsilon_{\lambda}}(p,q,\tau) = 
    < \epsilon_{\lambda}(t)^p \epsilon_{\lambda}(t+\tau)^q >
\label{eq17}
\end{equation} 
for moments of order $p>0$, $q>0$ and for $1 \le \tau \le \lambda$.
The stochastic integrals are split in order to separate the 
overlapping integration domains, corresponding to independent
random integrands. This gives:
\begin{eqnarray} 
   && C_{\epsilon_{\lambda}}(p,q,\tau)   =
   \lambda^{\frac{-\mu}{2}(p+q)} \times \nonumber \\
   && \,  \times <\exp p\sqrt{\mu}\int\limits_{t+1-\lambda}^{t+1+\tau-\lambda} (t+1-u)^{-1/2} dB(u)>
           \nonumber \\
    && \, \times <\exp q\sqrt{\mu}\int\limits_{t}^{t+\tau} (t+1+\tau-u)^{-1/2} dB(u)> \nonumber \\
     && \, \times <\exp \sqrt{\mu}\int\limits_{t+1+\tau-\lambda}^{t} 
         ( p(t+1-u)^{-1/2} + \nonumber \\
         && \, \, + q(t+1+\tau-u)^{-1/2} )dB(u)> \nonumber \\
	 & =&  I_1\lambda^{\frac{-\mu}{2}(p+q)}
      \exp \left(\frac{p^2 \mu}{2}\int\limits_{t+1-\lambda}^{t+1+\tau-\lambda} \frac{du}{t+1-u}
          \right) \times \nonumber \\
      && \times \exp \left( \frac{q^2 \mu}{2}\int\limits_{t}^{t+\tau} \frac{du}{t+1+\tau-u}
          \right) \nonumber\\
	 & = & I_1\lambda^{\frac{-\mu}{2}(p+q)}
	    \left( \frac{\lambda}{\lambda-\tau}\right)^{\frac{p^2\mu}{2}}
	    \left( \tau+1 \right)^{\frac{q^2\mu}{2}} 
\label{eq18}
\end{eqnarray} 
where $I_1$ is the last integral to evaluate:
\begin{eqnarray} 
  I_1 &=& \exp  \frac{\mu}{2}\int\limits_{t+1+\tau-\lambda}^{t} 
         ( p(t+1-u)^{-1/2} + \nonumber \\
         && + q(t+1+\tau-u)^{-1/2} )^2  \nonumber \\
	 & = & I_2 
	   \exp \left(\frac{p^2 \mu}{2}\int\limits_{t+1+\tau-\lambda}^{t} \frac{du}{t+1-u}
	      \right) \times \nonumber \\
	&& \times  \exp \left( \frac{q^2 \mu}{2}\int\limits_{t+1+\tau-\lambda}^{t} \frac{du}{t+1+\tau-u}
	     \right) \nonumber\\
	 & = & I_2
	    \left( \lambda-\tau \right)^{\frac{p^2\mu}{2}}
	    \left( \frac{\lambda}{\tau+1} \right)^{\frac{q^2\mu}{2}} 
\label{eq19}
\end{eqnarray} 
with finally
\begin{eqnarray} 
  I_2 &=& \exp\left( \mu pq \int\limits_{t+1+\tau-\lambda}^{t} 
         \frac{du}{ \sqrt{ (t+1-u)(t+1+\tau-u)}} \right) \nonumber \\
	 & = & \left( 
	   \frac{ \sqrt{\lambda-\tau}+\sqrt{\lambda}}
	        {1+\sqrt{\tau+1}}  \right)^{2\mu pq}
\label{eq20}
\end{eqnarray} 
where we have used the identity
\begin{equation} 
   \int \frac{dx}{\sqrt{x(x+a)}} = 2 \log \left( \sqrt{x}+\sqrt{x+a} \right)
 \label{eq21}
\end{equation} 
Eqs.(\ref{eq18}), (\ref{eq19}) and (\ref{eq20}) give:
\begin{equation} 
  C_{\epsilon_{\lambda}}(p,q,\tau)  = \lambda^{K(p)+K(q)}
  \left( \frac{ \sqrt{\lambda -\tau} + \sqrt{\lambda}} { 1+ \sqrt{\tau+1}}
  \right)^{k(p,q)}
\label{eq21b}
\end{equation} 
with $k(p,q)=2\left( K(p+q)-K(p)-K(q) \right)$.
Whenever $1 \ll \tau \ll \lambda$ this gives finally:
\begin{equation} 
  C_{\epsilon_{\lambda}}(p,q,\tau) \approx \lambda^{K(p+q)}
  {\tau}^{K(p)+K(q) -K(p+q)}
\label{eq22}
\end{equation} 
having the same asymptotic form as the generalized correlations for
discrete cascades (Eq. (\ref{eq6b})).

\indent On the other hand, we may introduce the logarithm, called here singularity process
\begin{eqnarray} 
  \gamma_{\lambda}(t) &=& \log \epsilon_{\lambda}(t) = 
  -\frac{\mu}{2} \log\lambda + \nonumber \\
  &+&  
     \mu^{1/2}\int\limits_{t+1-\lambda}^t (t+1-u)^{-1/2} dB(u) 
\label{eq23}
\end{eqnarray} 
and estimate its autocorrelation function for $ 1 \le \tau \le \lambda-1$ (using Appendix B):
\begin{eqnarray} 
   && C_{\gamma_{\lambda}}( \tau)  = < \gamma_{\lambda}(t)\gamma_{\lambda}(t+\tau)> 
   =  \left( \frac{\mu}{2} \log \lambda \right)^2  + \nonumber \\
      &&    \, +\mu <\int\limits_{t+1-\lambda}^t (t+1-u)^{-1/2}dB(u) \times \nonumber \\
       && \, \times   \int\limits_{t+1-\lambda+\tau}^{t+\tau} (t+1+\tau-u)^{-1/2}dB(u)>
\label{eq24}
\end{eqnarray} 
As before, the stochastic integral can be split into two non-overlapping
domains. Then, using Eq.(\ref{eqa12}) from Appendix B, we have:
\begin{eqnarray} 
  C_{\gamma_{\lambda}}( \tau)  &=&   \left( \frac{\mu}{2} \log \lambda \right)^2
      + \mu \int\limits_{t+1-\lambda+\tau}^t (t+1-u)^{-1/2} \times \nonumber \\
       && \times (t+1+\tau-u)^{-1/2}du
\label{eq25}
\end{eqnarray} 
and using again Eq.(\ref{eq21}), this gives:
\begin{eqnarray} 
  C_{\gamma_{\lambda}}( \tau)  & = &   \left( \frac{\mu}{2} \log \lambda \right)^2
      + 2\mu \log \frac{ \sqrt{\lambda-\tau}+\sqrt{\lambda}}{1+\sqrt{\tau+1}}
      \nonumber \\
      & \approx & A_{\lambda} - \mu \log \tau
\label{eq26}
\end{eqnarray} 
where $A_{\lambda}$ is a constant depending weakly on $\lambda$. The last line
used as before the assumption $1 \ll \tau \ll \lambda$. Equation (\ref{eq26}) is
very close to Eq. (\ref{eq9}) which was obtained for discrete cascades. 

\indent The stochastic process introduced in Eq. (\ref{eq14}) has thus all the 
characteristic properties of multifractal cascades: multiscaling of the
1-point statistics, together with long-range power-law correlations for
the process and log-corre\-lations for the logarithm of the process.
We consider below other properties of this equation: composition rules
 and differential forms.

\subsubsection{Differential properties}

Let us write stochastic equations for the processes
$\epsilon_{\lambda}(t)$ and $\gamma_{\lambda}(t)$. First, let us consider the
process $V_1(t)=\int_0^t (t+1-u)^{-1/2}dB(u)$. Applying the result given
in Appendix B, $V(t)$ is a martingale process (with respect to the natural filtration)
 having the differential
form:
\begin{equation} 
  \left\lbrace
    \begin{array}{l}
      dV_1(t) = dB(t)+W_1(t)dt \\
      W_1(t)=-\frac{1}{2} \int_0^t (t+1-u)^{-3/2}dB(u)
     \end{array}
   \right.
\label{eq27}
\end{equation} 
The same formula applied to the process $V_2(t)=\int_0^t (t+\lambda-u)^{-1/2}dB(u)$
gives
\begin{equation} 
  \left\lbrace
    \begin{array}{l}
      dV_2(t) = \lambda^{-1/2} dB(t)+W_2(t)dt \\
      W_2(t)=-\frac{1}{2} \int_0^t (t+\lambda-u)^{-3/2}dB(u)
     \end{array}
   \right.
\label{eq28}
\end{equation} 
These relations give then the stochastic process $\gamma_{\lambda}(t)$,
having as differential form $d \gamma_{\lambda}(t)=\mu^{1/2}(dV_1(t)-dV_2(t+1-\lambda))$:
\begin{equation} 
  \left\lbrace
    \begin{array}{l}
      d\gamma_{\lambda}(t) = \mu^{1/2} \left( dB(t) -\lambda^{-1/2}
         dB(t+1-\lambda) \right) -\frac{1}{2} W(t)dt  \\
      W(t)= \mu^{1/2} \int\limits_{t+1-\lambda}^t (t+1-u)^{-3/2}dB(u)
     \end{array}
   \right.
\label{eq29}
\end{equation} 
For large scale ratios ($\lambda \gg 1$), the term $\lambda^{-1/2}dB(t+1-\lambda)$ may be
neglected, since it is much smaller than $dB(t)$, so that the differential takes a simple form.

\indent We consider now the stochastic differential equation for $\epsilon_{\lambda}(t)
=\exp \gamma_{\lambda}(t)$. Ito stochastic calculus gives (recall that
$dX(t) = X(t)dY(t)+\frac{1}{2}X(t)dt$ for $X(t)=\exp Y(t)$):
\begin{equation} 
  d \epsilon_{\lambda}(t) = \mu^{1/2} \epsilon_{\lambda}(t) 
     \left( dB(t) +\frac{1}{2} (1-W(t)) dt \right)
\label{eq30}
\end{equation} 
where we have still neglected the term $\lambda^{-1/2}dB(t+1-\lambda)$,
and $W(t)$ is given in Eq.(\ref{eq29}). This provides the stochastic differential
equation having as a solution the multifractal lognormal process studied here.
It corresponds to a multiplicative process with a stochastic drift term.
This may be compared to other families of stochatic differential equations; it
can also be used for theoretical analyses, or to estimate (theoretically or numerically)
the predictability properties of such processes. Indeed the long-range 
correlation properties of such process correspond to a long memory that
can be exploited to optimize its predictability. This was already
done in another context \cite{mars96} and could be pushed further
for the present stochastic process. We do not develop further this point here,
and let such analysis for future studies.

\subsection{Scale statistics of the log-normal stochastic process}

In the previous section, we have considered the time
properties of the stochastic evolution equation,
for a given (and large) value of the scale ratio $\lambda$:
time correlations and time stochastic differential equation.
In this section we consider scale properties, assuming a
fixed time $t$ and considering the development of a cascade
at a fixed point, when $\lambda$ increases. We then define
\begin{equation} 
  E_t (\lambda) = \epsilon_{\lambda}(t)
\label{eq30b}
\end{equation} 
We take $\lambda$ as a continuous variable; in some instances
we consider the $\log$ of the scale ratio and take
as scale variable $\rho = \log \lambda$. 

\subsubsection{Semigroup composition properties}
Let us first recall that, in consequence of
their construction, discrete cascade, when considered
at a fixed position, can be considered as a discrete
Markov process in scale (see e. g. \cite{clev00}): since
successive multiplicative weights are independent,
the product of $n+m$ weights can be decomposed
into the product of $n$ weights times the product
of $m$ weights, possessing a semi-group property.

 \indent  We consider here a semigroup
property hidden in our stochastic
evolution equation.
\begin{eqnarray} 
  && E_t(\lambda_1\lambda_2) = (\lambda_1\lambda_2)^{-\mu/2} \times \nonumber \\
  && \times  \exp \left(  \mu^{1/2}\int\limits_{t+1-\lambda_1\lambda_2}^{t+1-\lambda_1} 
    (t+1-u)^{-1/2} dB(u) \right) \nonumber \\
    && \, \times
    \exp \left(  \mu^{1/2}\int\limits_{t+1-\lambda_1}^{t} 
    (t+1-u)^{-1/2} dB(u) \right) \nonumber \\
    && \stackrel{d}{=}  E_t(\lambda_1)\lambda_2^{-\mu/2} \times \nonumber \\
     && \times \exp \left(  \mu^{1/2}\int\limits_{t/\lambda_1-\lambda_2+1}^{t/\lambda_1} 
    {\lambda_1}^{1/2} \left( t+\lambda_1(1-v)\right)^{-1/2} dB(v) \right)  \nonumber \\
    && \stackrel{d}{=}  E_t(\lambda_1)
      \exp \left(  \mu^{1/2}\int\limits_{t'+1-\lambda_2}^{t'} 
    (t'+1-v)^{-1/2} dB(v) \right)  \nonumber \\
   & & \stackrel{d}{=}  E_t(\lambda_1) E_{\frac{t}{\lambda_1}}(\lambda_2)
\label{eq33}
\end{eqnarray} 
\noindent In the first equation the integration domain has been split in two
non-overlapping domains, so that the corresponding random variables are
independent. In the second equation, we have introduced a change of variable
($v-1=\frac{u-1}{\lambda_1}$)
and the fact that $dB(\lambda v) \stackrel{d}{=} \lambda^{1/2} dB(v)$.
The third and fourth equations are simple factorizations (with $t'=\frac{t}{\lambda_1}$)
using the definition
of $E_t(\lambda)$. As a result, Eq.(\ref{eq33}) is a decomposition property
``in distribution'', for a fixed time.

Let us note that usually in papers studying Markov properties of the
cascade in scale, the time or space variable do not appear. This is justified
by the fact that the process is stationary, and that one often considers the
probability instead of the process itself. To see this, let us introduce 
$\pi_{\rho}(x) = \Pr (\log E_t(e^{\rho})=x)$. Then taking the log of Eq.(\ref{eq33}), one
obtains an equality in distribution for the sum of random variables, leading to
a convolution of their probabilities:
\begin{equation} 
  \pi_{\rho_1+\rho_2} =
     \pi_{\rho_1} \otimes \pi_{\rho_2}
\label{eq31}
\end{equation} 
where $\rho_i = \log \lambda_i$. 

This recovers the convolution semi-group property whi\-ch can be found also in 
several recent papers \cite{male00,cast96,cast95}. 
In these papers, the convolution property has been shown to be directly linked
to a decomposition property (originally introduced in \cite{cast90}) of the pdf of the cascade process. 
This can also be obtained considering that 
the process $f(\rho)=\log E_t(e^{\rho})$ is a continuous Markov process with 
homogeneous transition
probability of the form $ Q_r(x,y) = \Pr (f(\rho+r)=y | f(\rho)=x)=q_r(y/x)$.
Then Chapman-Kolmogorov equation for transition probabilities can
be written on the following form

\begin{equation} 
  q_{r+s}(x) = \int q_r \left( \frac{x}{a} \right) q_s(a) da ;\,\,\; a>0
\label{eq35}
\end{equation} 

\noindent This equation is a general property of log-infinitely
divisible (log-ID) processes, but since characteristic functions have simple
expressions for such processes, it is often more convenient (as we do in the present paper) to 
characterize them using characteristic functions instead of
probability distributions.

\indent In a series of recent papers, Friedrich, Peinke and collaborators
\cite{frie97,naer97,renn01}
have experimentally shown that, for a fixed position, the velocity or dissipation
statistics in turbulence are Markov processes. 
Their Fokker-Planck approach can be shown to be linked
(in case of lognormal cascades) to Castaing's semi-group
decomposition approach \cite{cast97,ambl99,male00}. Below we consider in
this framework different properties in scale, leading to a Langevin
equation.

\subsubsection{Differential properties and a Langevin equation}

The series of papers about experimental evidence of a Markov
process for the energy cascade in turbulence have considered either
the velocity field or the dissipation field, and focused on a
Fokker-Planck or Langevin (differential) equations (for a link
between log-normal multifractal statistics and Fokker-Planck approach,
see \cite{cast97}). Here we give the
theoretical expression obtained for our process and compare our
result with experimental results published for the dissipation field
in turbulence.

A stochastic (Langevin) equation has been obtained by
Marcq and Naert \cite{marc99} for the logarithm of the cascade process
(more precisely the centered variable),
involving the logarithmic scale ratio variable $\rho = \log \lambda$.
We check here what this gives for our process. First for its 
generator, we have:

\begin{equation} 
  \left\lbrace
    \begin{array}{l}
      \gamma_t(\rho) = \log\epsilon_{\lambda}(t) = -\frac{\mu}{2} \rho
     + \mu^{1/2} G(e^{\rho}) \\
     G(x) = \int\limits_{t+1-x}^{t} (t+1-u)^{-1/2} dB(u) \\
     \end{array}
   \right.
\label{eq44}
\end{equation} 

\noindent This is easily differentiated and provides:
\begin{equation} 
  \left\lbrace
    \begin{array}{l}
      d\gamma_t(\rho) =  -\frac{\mu}{2} d\rho + \mu^{1/2}e^{\rho} dG(e^{\rho}) \\
      dG(x) = x^{-1/2}dB(t+1-x)  \\
     \end{array}
   \right.
\label{eq45}
\end{equation} 

\noindent This may be more simply written introducing another Brownian
motion $B_0(\lambda)=-B(t+1-\lambda)$ (recall that the time is frozen here),
and writing the equation for the centered singularity $\gamma_0=\gamma_t-<\gamma_t>$:
\begin{equation} 
  d\gamma_0(\rho) =  \sqrt{ \mu e^{\rho}} dB_0(e^{\rho})
\label{eq46}
\end{equation} 
where we used the property $<G(x)>=0$ and hence $<\gamma_t(\rho)>= -\frac{\mu}{2} \rho$.
For the process $E_t(\lambda)$ itself, this gives the following 
equation:
\begin{equation} 
  d E_t(\lambda) =  E_t(\lambda) \left(
    \frac{d \lambda}{2} (1-\frac{\mu}{\lambda}) + \mu^{1/2}\lambda^{1/2} dB_0(\lambda) \right)
\label{eq47}
\end{equation} 
This gives the Langevin equation in scale for our continuous 
multifractal process (for a fixed time).

\indent Our results can be compared to previous studies, some of them done in purely
experimental grounds. Let us first note that in some cases
(see \cite{naer97}) the study is done considering a Fokker-Planck
equation with drift and diffusion coefficients $D_1$ and $D_2$.
The associated stochastic differential equation (Langevin form)
is:
\begin{equation} 
  d \gamma(\rho) = D_1(\rho) d\rho +\sqrt{ 2 D_2(\rho)} dB(\rho)
\label{eq47b}
\end{equation} 

\noindent so that our results given in Eq.(\ref{eq46}) for the centered variable 
provide an equation analogous to Eq.(\ref{eq47b}) with:
\begin{equation} 
  \left\lbrace
    \begin{array}{l}
      D_1(\rho) =  0 \\
     D_2(\rho) = \frac{1}{2}\mu e^{\rho} 
     \end{array}
   \right.
\label{eq47c}
\end{equation} 
Our stochastic equation is thus analogous to a zero drift and and scale-dependent
diffusion coefficient, but with a subordinated Brownian motion
$B_0(e^{\rho})$, instead of the process itself.
Let us underline that this applies to a bare cascade, developing
from large to small scales, at a fixed time. This can be compared to the results
given by Cleve and Greiner \cite{clev00} for a discrete bare cascade, where they
also find a zero drift coefficient. 
On the other hand, experimental results on the Markov property of the
cascade correspond to properties of the dressed cascade and 
therefore may not be related directly to properties of the bare cascade.
This difference was already underlined in \cite{clev00} where it was
shown numerically (from a discrete cascade simulation)
that the dressing introduces a non-zero drift term.
The theoretical reasons to obtain a Markov property for dressed
quantities is also less straightforward than for bare ones. 
Nevertheless, we may compare the values given by our
bare model with dressed experimental results.
Naert et al. \cite{naer97} have experimentally obtained a non-zero drift value
and a constant $D_2$; later Marcq and Naert \cite{marc99} have found
a scale-dependent diffusion term of the form $D_2(\rho)
= D_0 e^{2 \delta \rho}$ with $D_0 \approx 0.01$ and
$\delta \approx 0.4$. This is closer to our result: $D_0$ is
clearly too small in   \cite{marc99}  compared to our expression
(recall that  experimentally, 
$\mu$ is close to $0.25$, leading to $D_0 \approx 0.12$) but the $\rho$-dependence
is almost the same since we would have $\delta=1/2$.
In any case we do not expect here to recover the same values, since
(i) there may be a bare-dressed difference in the coefficients and
(ii) the discrepancy between our theoretical expression
and experimental results are also likely to come from the fact that
we took for simplicity (as a first step) a log-normal cascade here, 
whereas the real picture in turbulence seems to be more
involved  (see e.g. \cite{schm92}).

Let us underline once more that this scale description is valid only
for a fixed position in time or in space, since the long-range
 correlations of turbulence in time or in space are not
 compatible with a Markov process. A more complete representation
 of the cascade process is then -- as we do in the present paper --
 to provide a time-scale description. In this time-scale space,
 the process involved is more complex than being simply 
 Markovian.

\subsubsection{Interpretation of multiplicative cascades as stochastic continuous products}

Discrete cascades correspond to a finite product of random variables.
For convenience reasons, the densification has been interpreted above as the exponential of
a stochastic integral. In fact, the natural way to densify the cascade
would be to introduce more and more intermediate scales in the finite
product, while keeping the total scale ratio fixed.
This corresponds to a continuous product (for deterministic
continuous products see \cite{doll79}).
Stochastic continuous products (also sometimes called stochastic
multiplicative integrals) have been introduced by McKean \cite{mcke60},
and developed by e.g.  Ibero \cite{iber76}, Doleans-Dade \cite{dole76} 
and Emery \cite{emer78} in the
framework of continuous martingales.
As first argued in \cite{schm01a}, continuous products are necessary
when one needs to introduce zeroes (such as in a continuous
generalization of the discrete $\beta$-model \cite{fris78}),
or when generalizing the 1D approach to matrix or tensorial cascades.
Stochastic continuous products provide in fact an interesting mathematical 
framework to study continuous cascades, since some results obtained in
this field may be applied to multifractal cascade processes.
Some first results in this approach  are given below.

\indent  Among the results obtained by Ibero \cite{iber76}
is the following: under general conditions on the square matrices functions
$f$ and $e$ (they can be
stochastic functions), the following
multiplicative stochastic integral (denoted as $\bigcap$, see Appendix C for a definition) is converging 
\begin{equation} 
 z(t) = \bigcap_0^t \exp \left( f(s)ds + e(s)dB(s) \right)
\label{eq48}
\end{equation} 
and is the solution of the following stochastic differential equation:
\begin{equation} 
 dz(t) = z(t) \left( (f(t)+\frac{1}{2}e^2(t))dt + e(t)dB(t) \right)
\label{eq49}
\end{equation} 
in the algebra of square matrices of order $n$ ($B$ is a matrix Brownian motion). 
This framework is useful for cascades of matrices,
which can be used to consider vectorial multifractal processes.
In this line, the generalization of  Eq.(\ref{eq14}), is, introducing continuous product:
\begin{equation} 
  \epsilon_{\lambda}(t) = \lambda^{-\mu/2} \bigcap_{t+1-\lambda}^t 
   \exp \left( \mu^{1/2} (t+1-u)^{-1/2} dB(u) \right)
\label{eq50}
\end{equation} 
which will give an equation analogous to  Eq. (\ref{eq30}) 
for square matrices. Indeed, for 1D processes, this equation
gives back Eq.(\ref{eq14}), while for the general multi-dimensional
case where $B$ is a matrix Brownian motion, this stochastic continuous
product can be shown to correspond to a matrix multifractal process.
This means that each element of such matrix is a 1D multifractal
process, while they also possess two-by-two scaling
properties. We do not explore further here the multifractal
and correlation properties of Eq. (\ref{eq50}); this will be done elsewhere.
Neverthelsess, it is clear that it would be interesting to develop such theoretical
framework in order to perform data analysis, such as for example
for analyzing multidimensional data series.
We discuss here a potential application of such
approach: these cascades of matrices may be used to model intermittent
anisotropic tensors in turbulence, such as the dissipation
rate $\epsilon_{ij}(t)$ where it is defined as
\begin{equation} 
   \epsilon_{ij}= 2 \nu < \frac{ \partial u_i}{\partial  x_k}
   \frac{ \partial u_j}{\partial  x_k}>
\label{eq51}
\end{equation} 
Indeed, according to Kolmogorov's 1941 hypothesis, for high
Reynolds number flows the cascade process is assumed to
``wash'' the details of the large-scale flow so that
after enough cascade steps the dissipation tensor becomes
isotropic. In fact experimental results in
homogeneous turbulence indicate that this may not be true
\cite{tavo81}.  
Several deterministic closure models have been developed 
to take into account this anisotropy for homogeneous and
non-homogeneous turbulence (see e.g. \cite{anto94,spez97}).
This type of model are fully
tensorial and can be numerically solved to provide
the prediction of some statistical quantities (such as
mean and second moments of the velocity field). But they need
too drastic hypothesis (in particular they are not compatible
with the long-range correlation properties of turbulence)
and they provide, at best, adequate predictions for moments up
to order 2 (see a general presentation on this
in \cite{pope00}). On the other hand, turbulent cascade
models are able to reproduce experimental fluctuations of the velocity field up
to moments of order 7 (see \cite{arne96}), but up to now they can deal
only with 1D processes. Because of this limitation, they have
offered until now very limited industrial applications.
If turbulent cascade models could be adapted to vectorial or tensorial
frameworks, it could, in a relatively short time, lead
to much more reliable predictions for industrial flows.
This promising research direction is not developed further here.

\begin{figure}
\begin{center}
\resizebox{0.49\textwidth}{!}{%  
  \includegraphics{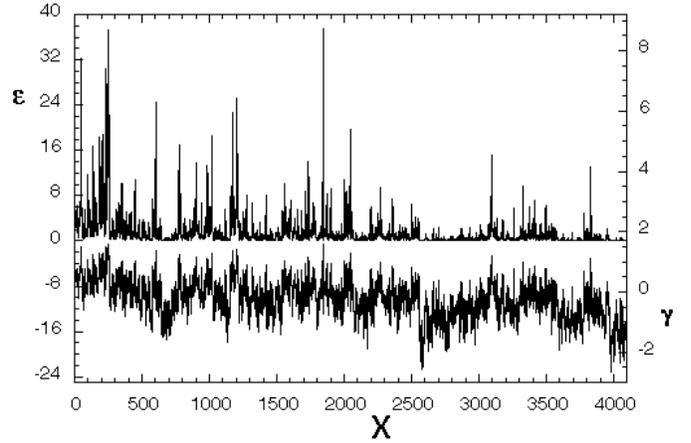}
}
\end{center}
\caption{A sample of a log-normal discrete cascade with total scale ratio 
$2^{12}$. The generator $\gamma$ (a correlated Gaussian noise) and the multifractal field
$\epsilon$ are shown.}
\label{fig:3}       % Give a unique label
\end{figure}

\begin{figure}
\begin{center}
\resizebox{0.40\textwidth}{!}{%  
  \includegraphics{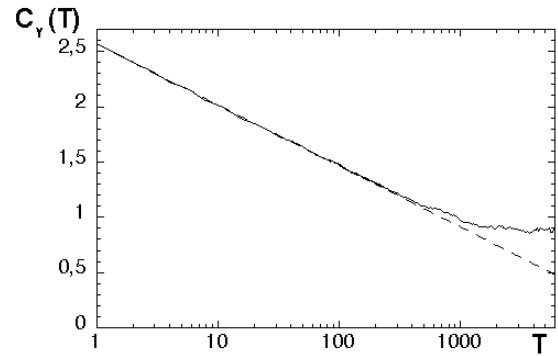}
}
\end{center}
\caption{The autocorrelation function of the generator, estimated from 
100 realizations of a discrete log-normal cascade with $\mu =0.25$. 
As expected this shows a logarithmic decrease over the whole scaling range. 
Dotted line: predicted logarithmic decay of the form $-\mu \log T$.}
\label{fig:4}       % Give a unique label
\end{figure}

\section{Simulations}
We perform here numerical simulations of discrete and continuous cascades. The aim of these 
simulations is to check the theoretical relation given in previous sections, and to
compare discrete and continuous cases.

\subsection{Simulation of discrete log-normal cascades}
We perform here a numerical simulation of a discrete log-normal cascade
with $\lambda_1 =2$ and weights given by Eq.(\ref{eq12}).
This is illustrated here with $n=12$ cascade steps, corresponding to 
a total scale ratio of $\lambda = 2^{12}=4096$, and a value of the
intermittency parameter close to experimental results for the dissipation:
$\mu =0.25$.  Figure 3 shows the resulting field for the cascade process
$\epsilon$ and the singularity $\gamma$. It is visible that $\epsilon$
is highly intermittent while $\gamma$ is a correlated noise. This 
correlation is illustrated in Fig. 4, showing $C_{\gamma}(\tau)$ versus $\log \tau$:
the logarithmic decay is extremely well verified for almost 3 orders of magnitude, with
a slope of about 0.24, close to the predicted value, which is $\sigma ^2/\log 2 =\mu$.
Figure 5 represents the power spectrum of $\gamma$ and of $\epsilon$ in log-log plot.
These are
both scaling, with respectively theoretical slopes of -1 and $-1+\mu$: see Eq. (\ref{eq10}) and
the Fourier transform of Eq. (\ref{eq6}). Experimental results are very close to
theoretical power laws for the whole range of scales considered.

\begin{figure}
\begin{center}
\resizebox{0.40\textwidth}{!}{%  
  \includegraphics{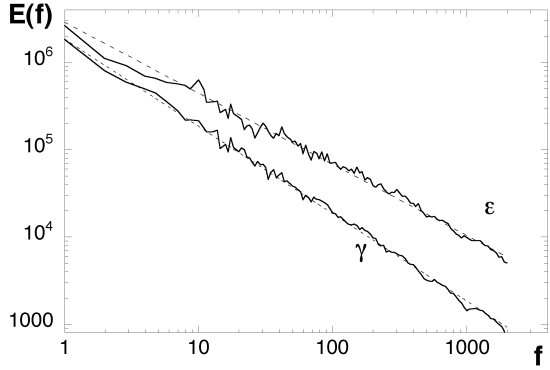}
}
\end{center}
\caption{The Fourier power spectrum of 100 realizations of a discrete log-normal 
multifractal field and of its generator. As expected, over the whole scaling range, the generator 
displays an exactly $f^{-1}$  power spectrum, and the multifractal field a  $f^{\mu-1}$. 
Dotted lines provide theoretical behaviours.}
\label{fig:5}       % Give a unique label
\end{figure}

\begin{figure}
\begin{center}
\resizebox{0.4\textwidth}{!}{%  
  \includegraphics{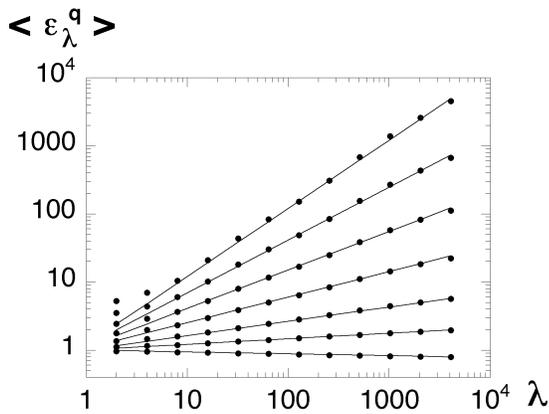}
}
\end{center}
\caption{The scaling of the moments of order 0.5, 1.5, 2, 2.5, 3, 3.5, and 4 (from below to above), 
estimated for 100 realizations of the discrete log-normal multifractal cascade. 
The different straight lines indicate the accuracy of the scaling property. 
The slopes of these straight lines give estimates of $K(q)$.}
\label{fig:6}       % Give a unique label
\end{figure}

Figure 6 is a direct test of the scaling of the 
resulting field, for various order of moments: 100 realizations (each of length
4096) have been
taken into account for the estimation of statistical moments. The dressed moments
are represented in log-log plot, for different values of $q$; the straight lines
confirm a scaling law of the form given by Eq.\ref{eq4} for the dressed field.
The slope of these straight lines give experimental estimates of $K(q)$, and
the resulting curve is shown in Fig. 7, compared to the theoretical bare expression
$\frac{\mu}{2}(q^2-q)$. 
There is an excellent agreement until a critical
moment $q_s \approx 2.4$, above which the estimated function is linear: this corresponds
to a maximum order of moment that can be estimated with a finite number of samples, since
there are not enough realizations for an accurate estimation of the scaling exponent corresponding
to larger order of moments. This may also correspond to the critical moment above which
moments of the dressed field diverge; in this case experimental estimates are not
infinite, but the exponent $K(q)$ becomes linear (see \cite{schm94} for a discussion
on bare/dressed exponents in an experimental context). 
In any case, experimental analysis shows that the simulated field has scaling
moments, and as expected for low orders of moments, the moment function is close
to the theoretical bare curve.

\begin{figure}
\begin{center}
\resizebox{0.45\textwidth}{!}{%  
  \includegraphics{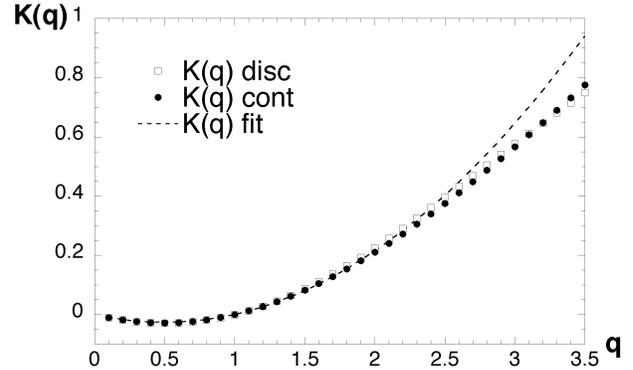}
}
\end{center}
\caption{The $K(q)$ function estimated from 100 realizations of the discrete 
multifractal simulation (open squares), from 106 points of a continuous simulation 
(filled dots), compared to the theoretical curve given by Equ. (17) (dotted line). 
There is an excellent agreement until a critical moment corresponding to the 
maximum moment that can be accurately estimated with a finite sampling.}
\label{fig:7}       % Give a unique label
\end{figure}

\begin{figure}
\begin{center}
\resizebox{0.49\textwidth}{!}{%  
  \includegraphics{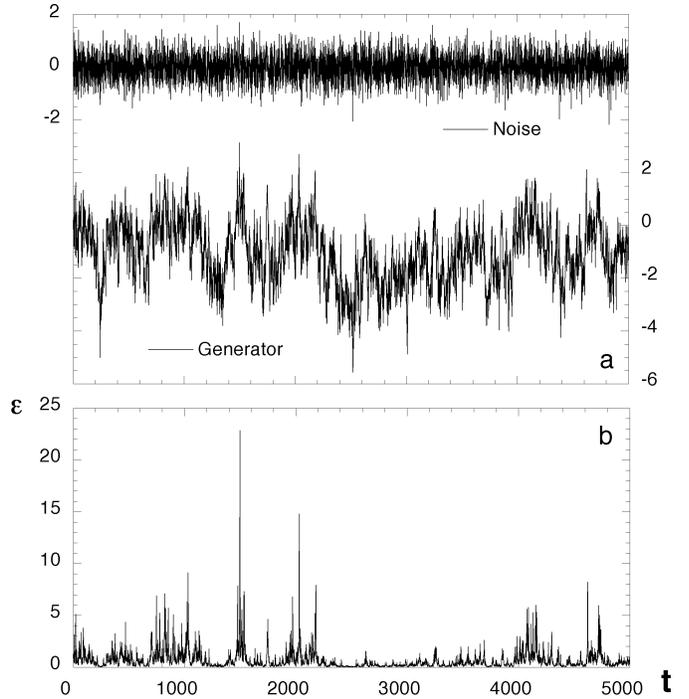}
}
\end{center}
\caption{(a) A sample of the Gaussian noise and of the generator; the 
generator is a moving average over the past of the noise, and corresponds
 to a correlated Gaussian noise. (b) A sample of the multifractal process, 
 corresponding to the exponential of the generator of Fig. (8a). }
\label{fig:8}       % Give a unique label
\end{figure}

\subsection{Simulation of continuous and causal log-normal processes}
Here we provide a stochastic simulation of the process given by
 Eq. (\ref{eq14}), choosing $\lambda = 5,000$ and $\mu =0.25$. 
 This is done sequentially, generating the noise and a moving average 
 over this noise. An arbitrary large number of points can be generated
 with this equation. We represent in Fig. 8a-b a sample path of 5,000 points
 for the process and its generator: Fig. 8a shows the Gaussian uncorrelated noise,
 together with the generator process, which is a moving average over this noise
 (with a specific kernel). This correlated noise has the same visual
 aspect as the generator obtained in Fig. 3 for discrete cascades.
 Figure 8b shows the field $\epsilon$ itself (the exponential of the generator): it presents
 as expected a huge intermittency.

\begin{figure}
\begin{center}
\resizebox{0.4\textwidth}{!}{%  
  \includegraphics{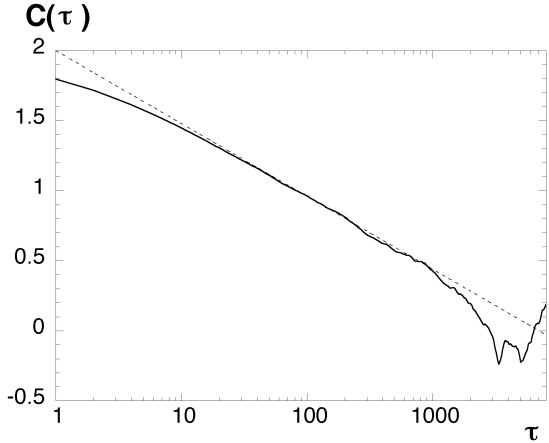}
}
\end{center}
\caption{The autocorrelation function of the generator, estimated from a 
continuous stochastic process with $\mu=0.25$. As expected this shows 
a logarithmic decrease for large time lags. Dotted line: predicted logarithmic decay 
of the form $-\mu \log T$ with an experimental $\mu$ value of 0.23.}
\label{fig:9}       % Give a unique label
\end{figure}

\begin{figure}
\begin{center}
\resizebox{0.4\textwidth}{!}{%  
  \includegraphics{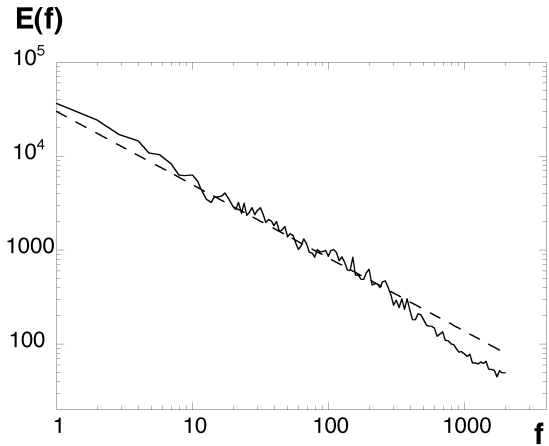}
}
\end{center}
\caption{The Fourier power spectrum of the continuous stochastic process. 
As expected, over the whole scaling range the multifractal field displays a  $f^{\mu-1}$ regime 
with an experimental value of $\mu=0.23$ (dotted line).}
\label{fig:10}       % Give a unique label
\end{figure}

 To estimate the statistics of the path, we have generated $10^6$ points.
 The scaling of their power spectra, together with their correlation properties, have been
 verified. The logarithmic decay of the autocorrelation function of the singularity process
 is shown in Fig. 9: it is nicely verified with a parameter of $\mu = 0.23$ close
 to the theoretical value. For short time lags (smaller than about 10) the experimental
 curve leaves the straight dotted line corresponding to the fit; this is expected since the 
 logarithmic decay was shown to be an asymptotic law for large enough time lags.
 Figure 10 shows the power spectrum of the field $\epsilon$ which is scaling
 with a fit of the form $f^{\mu-1}$ as expected, with an exponent of $-0.77$,
 close to the expected value.
 The resulting moment function $K(q)$ is also shown in Fig. 7, as a comparison with the theoretical
 expression and the function estimated for the discrete simulation. The value
 corresponding to the continuous simulation is very close to previous results for
 the discrete simulation, and the same comments about bare/dressed scaling exponents
 apply.
 These different results confirm the different asymptotic expressions
 given for $1 \ll \tau \ll \lambda$ and the multifractal properties of the stochastic simulation.

\section{Concluding comments}
We have explored here the statistical properties of time series
generated by a continuous and causal stochastic equation
taking the form of the exponential of a stochastic integral.
We have first shown that such generating equation indeed 
corresponds to a log-normal multifractal: this was verified with
one point and two points statistics, either analytically and also
numerically, in the asymptotic regime (for large enough time lags). We have also shown the composition rule of such
equation, and considered its differential properties. This 
equation depends on 2 parameters: the intermittency parameter
$\mu$ and the total scale ratio $\lambda$. For a fixed time and a 
variable scale ratio parameter, another type of process is 
obtained; the differential with respect to scale parameter leads
to a Langevin equation for the logarithm of the process. We have linked
this approach to experimental results about Langevin equations for
the dissipation in turbulence. Of course, this does not indicate that
log-normal cascades are the best models for turbulent intermittency.
This is rather the exploration of an important and classical multifractal
model; other continuous multifractals are associated to more complex
generating equations having less straightforward differential properties.

This equation for log-normal multifractals is nevertheless believed to
be of great interest since a quite simple stochastic equation generates
a causal and continuous multifractal process, that can be used as a 
rather realistic approximation for various applications. It is also the 
first explicit continuous equation of its kind. This approach is finally
believed to open the door to many interesting generalizations, including
cascades of matrices using a stochastic continuous product framework,
which is only quickly evoked here.

\section*{Acknowledgement}

We thank L. Coutin and D. Marsan for useful discussions, and 
the referees for their useful suggestions.

\section*{Appendix A. Characteristic functions and infinitely divisible distributions}
We recall here some useful properties of characteristic functions,
infinitely divisible distributions, Gaussian and L\'evy-stable
random variables.
Useful references for these topics are Feller \cite{fell71}, Janicki and Weron \cite{jani94},
and Samorodnistky and Taqqu \cite{samo94}.

Let $X$ be a random variable with probability density $p(x)$.
The usual characteristic function is the Fourier characteristic
function $\phi(k)$ defined as
\begin{equation} 
  \phi(k)=<e^{ikx}>
\label{eqa1}
\end{equation} 
Since when studying multiplicative cascades real moments are introduced,
we use here Laplace characteristic functions, assuming that the positive tail of
the probability density decreases fast enough for the integral to converge:
\begin{equation} 
  f(q)=<e^{qx}>=\int\limits_{-\infty}^{+\infty} e^{qx} p(x)dx
\label{eqa2}
\end{equation} 
One thus have $f(q)=\phi(-iq)$ where $\phi(q)$ are given 
in the usual tables of characteristic functions.

\indent Infinite divisibility is a property that has no simple
expression for probability densities. Instead, this property is very
simply characterized using characteristic functions.
A probability distribution is said to be infinitely divisible (ID) if its
characteristic function $f$ has the following property: for any $n$ integer,
there exist a characteristic function $f_n$ such that:
\begin{equation} 
  f_n^n = f
\label{eqa3}
\end{equation} 
In such a case one can introduce the second characteristic function $\Psi = \log f$:
\begin{equation} 
  \Psi(q) = \log <e^{qx}>
\label{eqa4}
\end{equation} 
Let us consider a second characteristic function $\Psi(q)$. If for any $a>0$, $a\Psi(q)$ is
still a second characteristic function, then the distribution associated to $\Psi(q)$ is
ID. 

\indent  Let us give two examples: a Poisson distribution with mean $\alpha$
has a second characteristic function of
\begin{equation} 
  \Psi(q) = \alpha (e^q -1)
\label{eqa5}
\end{equation} 
L\'evy-stable random variables can be defined as follows: let $(X_i)_{i=1..n}$ 
be independent random variables of the same law, and $n$ any integer. Then the
law is stable if there exists $a_n >0$ and $b_n$ such that
$a_n \sum_{i=1}^n X_i - b_n$ has the same law as the $X_i$.
Strictly stable random variables are the ones for which $b_n=0$, corresponding
to centered $<X_i>=0$ variables.
Using the second characteristic function, it is easily seen that L\'evy
random variables, for which
\begin{equation} 
  \Psi(q) = \Gamma^{\alpha} q^{\alpha}
\label{eqa6}
\end{equation} 
are stable with $a_n=n^{1/\alpha}$. The parameter $\Gamma>0$ is called
the scale or dispersion parameter; it plays the same role
as the variance for gaussian random variables: for $\alpha=2$,
one recovers gaussian random variables with $\sigma^2 = \Gamma^2/2$.
L\'evy-stable random variables are characterized by the basic
parameter $\alpha$, limited to the range $0 \le \alpha \le 2$.
One may note that L\'evy-stable variables can take high values with
hyperbolic probability tails, so that for the Laplace characteristic
function to converge, one must consider only skewed L\'evy
variables, for which the hyperbolic tail corresponds to
negative values only. In this case, the Laplace characteristic function
is defined for moments $q>0$.

\section*{Appendix B. Stable stochastic integrals}
A useful reference for this topic is e.g. Janicki and Weron \cite{jani94}.
The rescaling properties of stable random variables can be used
to define stable stochastic integrals.
Let us consider again $Y=\sum_{i=1}^n X_i$, where
$(X_i)_{i=1..n}$ are independent strictly stable 
random variables having the same law, with parameters $\alpha$ and
$\Gamma_x$. Then one has $\Psi_Y(q) = n \Psi_x(q)=n\Gamma_x^{\alpha}q^{\alpha}$
showing that:
\begin{equation} 
  \Gamma_y = n^{1/\alpha} \Gamma_x
\label{eqa7}
\end{equation} 
This property can be used to build stable random measures
and in a consistent way, stable stochastic integrals.

\indent  For an interval of width $dx$, $M(dx)$ is defined as a strictly
stable random measure, i.e. a strictly stable random variable of index
$\alpha$ and
of scale parameter $\Gamma_M = (dx)^{1/\alpha}$. It has the following second
characteristic function:
\begin{equation} 
  \Psi_M(q) = \left( \Gamma_M q \right)^{\alpha} =q^{\alpha} dx
\label{eqa8}
\end{equation} 
For a positive valued function $F$, such that $\int_a^b F^{\alpha}(x)dx$
exists, the stochastic integral $I=\int_a^b F(x)M(dx)$ is a random variable defined
as follows: it is  the strictly stable random variable of scale parameter
\begin{equation} 
  \Gamma_I = \left( \int_a^b F^{\alpha}(x)dx \right)^{1/\alpha}
\label{eqa9}
\end{equation}   
which can be rewritten in the following way, using its second Laplace characteristic function:
\begin{equation} 
  \log < e^{ q \int_a^b F(x)M(dx)}>
    = \left( \int_a^b F^{\alpha}(x)dx \right) q^{\alpha}
\label{eqa10}
\end{equation}   
This fully characterizes stable stochastic integrals. In particular, for
$\alpha=2$,  the
Gaussian stochastic integral $I=\int_a^b F(x)dB(x)$ is still a Gaussian
random variable with the variance
\begin{equation} 
  \sigma_I^2 = \int\limits_a^b F^2(x) dx
\label{eqa11}
\end{equation}   
Gaussian stochastic integrals have also the property:
\begin{equation} 
   < \int_{E_1}F(x)dB(x) \int_{E_2}G(x)dB(x)>
     = \int_{E_1 \cap E_2} F(x)G(x)dx 
\label{eqa12}
\end{equation}   
Indeed, if $E_1 \cap E_2 =\emptyset $ the two integrals and independent
random variables and whenever $E_1 \cup E_2 \ne \emptyset$ only
the intersection contributes to the correlation.

\indent Let us finally consider the stochastic process defined as:
\begin{equation} 
   W_K(t)= \int_0^t K(t,s) dB(s)
\label{eqa13}
\end{equation}   
where the kernel $K$ possesses some regularity conditions:
\begin{eqnarray}    
  &&   K(t,s)  =  0 ; \; \; \;s>t \\
   &&  \int_0^T \left( \int_0^u \partial_t K(u,s)^2 ds \right)^{1/2} + \nonumber \\
      && \;\;\;\; + \int_0^T  K^2 (s,s)ds < \infty
\label{eqa14}
\end{eqnarray} 
then $W_K(t)$ is a semimartingale with respect to the usual
filtration and has the following representation
\begin{equation} 
   dW_K(t)= K(t,t) dB(t) + W_{\partial_t K}(t) dt
\label{eqa15}
\end{equation}   
See for more details Carmona and Coutin \cite{carm00}.

\section*{Appendix C. Stochastic continuous product}
Stochastic multiplicative integrals or continuous products have been introduced
in the field of continuous martingales (see McKean \cite{mcke60},
and developed by e.g.  Ibero \cite{iber76}, Doleans-Dade \cite{dole76} , Emery \cite{emer78} 
and also Karandikar \cite{kara83}). Below we propose a personal and short didactic
introduction to such objects. 

Infinitely divisible distributions are naturally associated to
random measures, which can also be called random additive set function.
The discrete addition of ID random variables, when densified, 
leads to stochastic integrals. The same way, one wants here to
densify a discrete multiplication of random variables, to obtain
a continuous product. This can be done introducing 
random multiplicative set functions (RMSF). 
We associate with a measure $m(A)$ for a set $A \ne \emptyset$, a multiplicative set function
$n(A)$  defined as $n(A)=e^{m(A)}$.
Then for $A$ and $B$ such that $A \cap B = \emptyset$, one has
$n(A \cup B)=n(A)n(B)$.
A random multiplicative set function $N$ can then be defined for
a set  $A$: $N(A)=e^{M(A)}$ and is characterized by the control measure $m(A)$.
It is clear that for  $A$ and $B$ with a void intersection, one has multiplicative
random variables: $N(A \cup B)=N(A)N(B)$.

This is useful to introduce a random continuous product.
Let us take non-intersecting intervals $A_i$ and introduce the random variable $P$:
\begin{equation} 
   P = \prod_{i=1}^{n}N(A_i)
\label{eqC4}
\end{equation} 
This can be generalized to a continuous limit, taking intervals of
the form $A_i =  [a+i\frac{b-a}{n},a+(i+1)\frac{b-a}{n}]$,
whose reunion is the interval  $[a,b]$ and whose width decreases with  $n$. 
This way one has the limit:
\begin{equation} 
          {\bf \bigcap}_a^b N(dx) = 
   \lim_{n \to \infty} \prod_{i=0}^{n-1} 
N \left( [a+i\frac{b-a}{n},a+(i+1)\frac{b-a}{n}]\right)
\label{eqC5}
\end{equation} 
One obtains a stochastic continuous product, as the limit
of a product of random multiplicative set functions, noted  ${\bf \bigcap}$.
It has the following basic property:
\begin{equation} 
  < \left( {\bf \bigcap}_A N(dx)\right)^q > = e^{m(A)\Psi_0(q)}
 \label{eqC7}
 \end{equation} 
which can be used for continuous cascades ($\Psi_0(q)$ is a reference
second Laplace characteristic function).

%and subsection a unique label (see Sect.~\ref{sec:1}).
%%\begin{figure*}
%%\vspace*{5cm}       % Give the correct figure height in cm
%%\caption{Please write your figure caption here}
%%\label{fig:2}       % Give a unique label
%%\end{figure*}
%\begin{table}
%\begin{tabular}{lll}
%\hline\noalign{\smallskip}
%\end{tabular}
% Or use
%\vspace*{5cm}  % with the correct table height
%\end{table}
%
% BibTeX users please use
% \bibliographystyle{}
% \bibliography{}

\end{document}